\documentclass[twocolumn,showpacs,preprintnumbers,amsmath,amssymb,floatfix]{revtex4}

\usepackage{graphicx}
\usepackage{dcolumn}
\usepackage{bm}

\begin{document}

\preprint{APS/123-QED}

\title{Spreading on a complex network avoiding certain motifs}

\author{Tomas Alarcon}
\email{a.tomas@imperial.ac.uk}
\author{Henrik Jeldtoft Jensen}
\email{h.jensen@imperial.ac.uk}
\affiliation{Institute of Mathematical Sciences, 53 Princes' Gate, Imperial College London, London SW7 2PG \& Department of Mathematics, Imperial College London, South Kensington campus, London SW7 2AZ, UK}

\date{\today}

\begin{abstract}
 Spreading of either information or matter can often be treated as a network problem. It can be of great importance to be able to estimate the likelihood that spreading through a network reaches essentially the entire network while still not reaching certain sub-classes of the network. We show that excluding nodes and edges from the network has a subtle effect on the percolation. We study two specific examples of degree distributions (exponential and scale free) for which analytical solutions can be obtained. The two cases exhibit qualitatively different behavior.
\end{abstract}



\maketitle
When information, or matter, spreads through a network, the local dynamics will typically be related to the local community structure \cite{newman2006}. This may be because the quantity that flows has an effect on the state of the local sub-community, or it may be that information flowing through a network will trigger a particular type of response when it reaches certain topological local sub-net structures. One may think of resonance effects if the flow involves activation of dynamical variables on the nodes. Other examples can be found in the effect of unbalanced triangles in social networks \cite{antal2005} or frustration effects when the dynamics corresponds to optimization. The prototype of the latter example is the role plaid by frustrated loops, like {\it e.g.} triangles when optimizing an anti-ferromagnetic energy functional on a network \cite{fisher1991}. Similarly, in sociology or epidemiology it can be of great interest to know how likely it is that a macroscopic proportion of the population will be touched by a spreading quantity, while certain sub-populations remain untouched. In the same vein, the fact that information tends to get trapped within communities \cite{rosvall2008,lambiotte2008}, provides a rationale for, under the right circumstances, trying to avoid such communities. 

The general problem of estimating the probability that spreading on a network reaches a macroscopic part of the entire network, while a certain type of motifs remain untouched, is obviously very complicated in its full generality. To make some initial headway we consider the special problem concerning spreading on random networks and compute the probability that the flow percolate to a macroscopic fraction of the network while avoiding a certain fraction of triangular motifs. We are able to express this probability in terms of the degree distribution and the edge clustering coefficients.  

Our main objective is to analyze how exclusion of triangular motifs affects percolation on weakly-clustered networks, i.e. how the spreading manages to reach a macroscopic part of the network without touching a give subset of triangular motifs. Typically one would expect that removing edges and nodes will make it harder for a process to spread across the network. This we confirm. However, the scenario is complex. For networks with an exponential degree distribution two types of behavior exist. For high edge-multiplicity the average degree, for which the spreading percolates on the decimated  network, increases as function of the number of removed triangles. In contrast, for low multiplicity the onset of percolation depends in a non-monotonous way on the number of excluded triangles. Scale free networks in contrast  only exhibit  the monotonous behavior.

{\it Model} -- We now consider a network characterized by the following two quantities: the degree distribution $P(k)$ and the nodal clustering coefficient $c(k)$. In the limit of weak clustering, we are able to express the conditions for percolation without hitting the designated triangles in terms of these two quantities. The procedure of our computation consists of removing, by random sampling, a certain proportion $T<1$ of all the triangular motifs of the original network. The statistical characteristics of the sampled network are expressed in terms of the distributions for the original network. Next the percolation process on the sampled network is studied and the percolation threshold calculated. This calculation allows us to determine under which conditions the spreading process percolates on the original network while leaving at least a fraction $1-T$ triangular motifs untouched. The approach we use consists of removing those nodes (and the corresponding edges) that belong to the proportion $T$ of triangles within the original network. The network generated by the remaining nodes and edges (the sampled network) consists thus of the proportion of the network that does not include any of the designated ``no-go'' motifs. The analysis of percolation on the sampled network allows us to study the spreading processes on the original network which stay clear of certain communities. In order to advance this program, we need to calculate how the relevant quantities on the sampled network relate to the original one. In other words, we need to parametrize the sampled network in terms of $P(k)$ and $c(k)$. Stumpf \& Wiuf have extensively analyzed the properties of binomial sampling of complex networks in relation to the validity of the inference of properties of the entire network from those of a (randomly sampled) sub-network \cite{stumpf2005,wiuf2006}. These authors have dealt mostly with the effect of binomial sampling on the degree distribution. Here, we extend the analysis to motif sampling and also consider the effect of sampling on the nodal clustering coefficient. We present only a brief summary of our main results. An extended treatment including the derivations in full detail will be presented somewhere else \cite{alarcon2009}.

\begin{figure*}
$\begin{array}{cc}
\mbox{Exponential} & \mbox{Scale-free}\\
\includegraphics[scale=0.2]{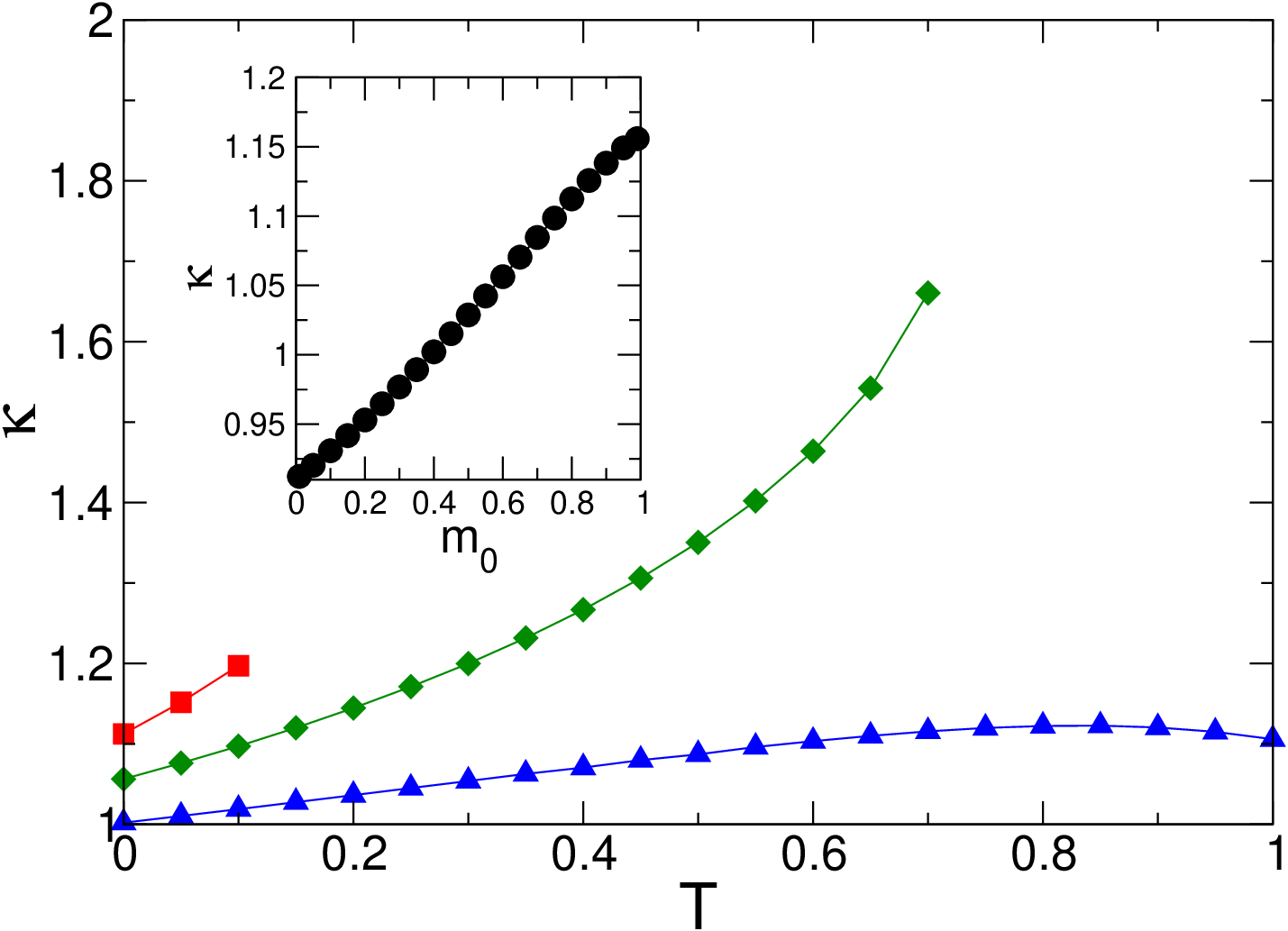} & \includegraphics[scale=0.2]{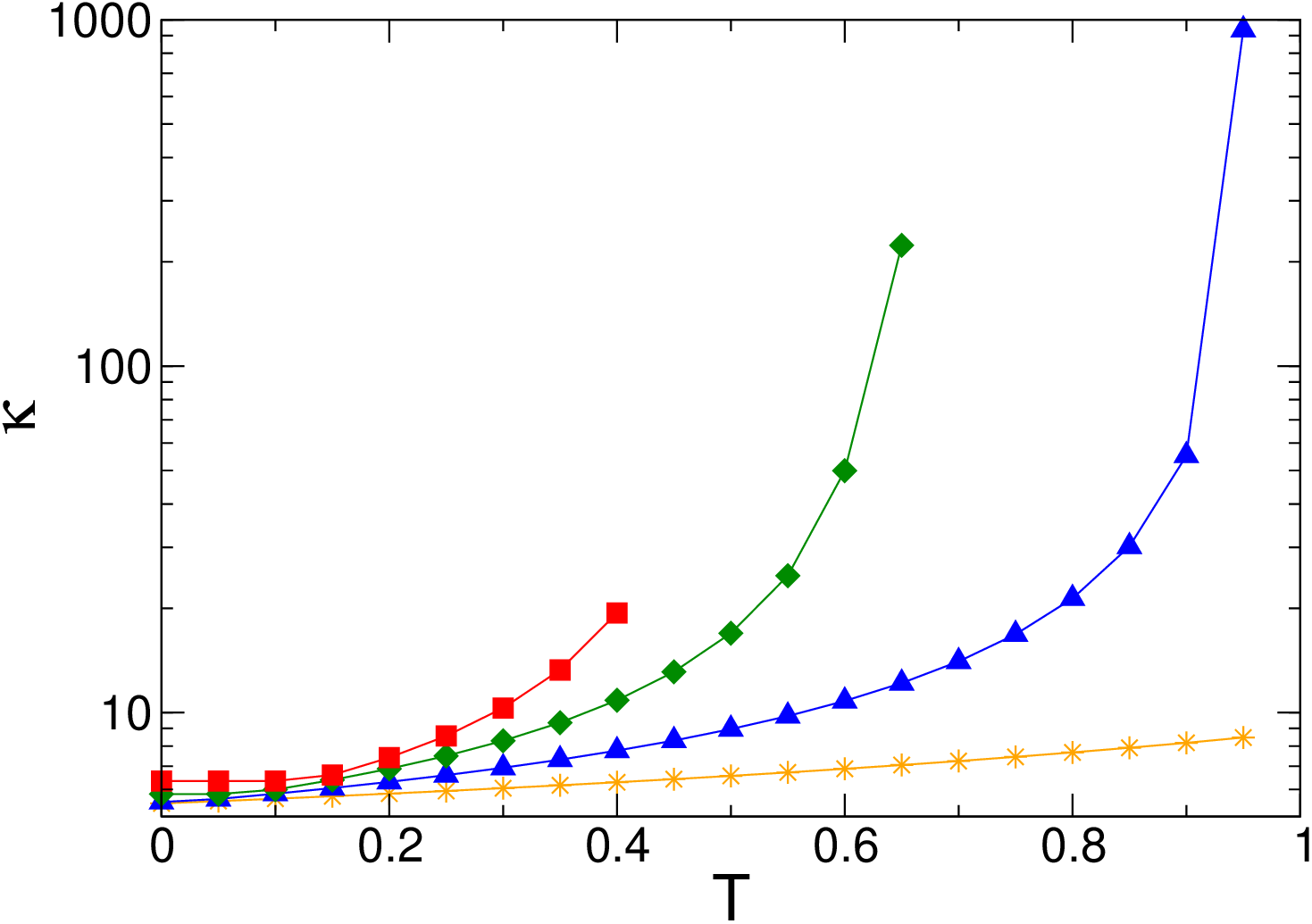}
\end{array}$
\caption{The analytic result for the percolation threshold $\kappa$ (corresponding to the critical value of the cut-off parameter $\beta$ in Eqs. (\ref{eq:exp1}) and (\ref{eq:sf1})) for networks with exponential (left) and scale-free (right) degree distribution. Comparing the two panels highlights the different response of between these two types of networks to the sampling process: whereas exponential networks show two different types of behavior depending on the edge multiplicity of the original network (i.e. monotonous dependence of $\kappa$ on $T$ for large clustering as opposed to non-monotonous dependence for lower clustering), scale-free networks exhibit always the same behavior (i.e. percolation being hindered by sampling) regardless of the edge multiplicity of the original network. Color code: black lines (inset) correspond to the unsampled exponential network for different values of the average multiplicity $m$, orange lines correspond to $m=0.2$ (only shown for scale-free networks), blue lines to $m=0.4$ for different values of $T$, green lines correspond to $m=0.6$, and red lines correspond to $m=0.8$.\label{fig:5}}
\end{figure*}

We briefly summarize the effect of sampling on the degree distribution as analyzed by Stumpf \& Wiuf\cite{stumpf2005}.  Define the quantity $p=\frac{1}{\langle k \rangle}\sum_{k}kQ(k)P(k)$ where $Q(k)$ is the probability of sampling a node with degree $k$. The quantity $p$ is the entry to the binomial distribution that determines whether a node is removed or not. Accordingly, within the sampled network a node has degree $l$ with probability $P_S(l)$ given by:

\begin{equation}\label{eq:samp2}
P_S(l)=\sum_{k\geq l}^{\infty}\left(\begin{array}{c}k\\l\end{array}\right)\pi(k)p^l(1-p)^{k-l}
\end{equation}

\noindent where $\pi(k)$ is given by: $\pi(k)=Q(k)P(k)/\langle Q \rangle$ with $\langle Q \rangle=\sum_{k}Q(k)P(k)$ is the total weight relative to the full network actually sampled.  The generating function corresponding to the sampled network, $G_S(x)$, is given by:

\begin{equation}\label{eq:samp4}
\nonumber G_S(x)= \sum_{k=0}\pi(k)(1-p+px)^k=G_{\pi}(1-p+px)\\
\end{equation}

Let us now assume that a random fraction $T<1$ of all the triangles within the network are chosen and designated as the motifs to avoid. The probability of a node being sampled is given by $Q(k)=\left(1-Tc(k)\right)^{k(k-1)/2}$, where $c(k)$ is the nodal clustering coefficient. This quantity can be interpreted as the probability that a node of degree $k$ belongs to a triangle \cite{serrano2006a}. 

The analytical approach we develop below assumes the  \emph{weak clustering condition}, i.e. the average edge multiplicity $m_0\leq 1$. Serrano \& Bogu\~n\'a \cite{serrano2006a} have argued that this condition can be expressed in terms of the clustering coefficient as $c(k)\lesssim c_0/(k-1)$. We will assume that $c(k)=c_0/(k-1)^{\alpha}$ with $\alpha\geq 2$. Under these conditions, an expansion in powers of $Tc(k)$ of $Q(k)$ can be performed. To first order, we obtain:

\begin{equation}\label{eq:samp6}
Q(k)\simeq 1-T\frac{k(k-1)}{2}c(k)
\end{equation}

We now turn to the analysis of the effect of sampling on different measures of clustering, in particular we consider the nodal clustering coefficient, $c(k)$, and the edge clustering coefficient $c(k,k^{\prime})$. Hereafter, we will assume the original network to be uncorrelated and therefore we will be limited to study weakly clustered networks \cite{serrano2006a}.

Triangular motifs are randomly removed in a uniform manner, independently of the degrees of the nodes composing the triangle. Therefore, the number of triangles within the degree class $k$ within the sampled network, ${\cal T}_S(k)$, is given in terms of the corresponding number of triangles in the original network, ${\cal T}(k)$: ${\cal T}_S(k)=(1-T){\cal T}(k)$. Taking into account that ${\cal T}(k)$ and $c(k)$ must be such that ${\cal T}(k)=\frac{1}{2}NP(k)k(k-1)c(k)$, Eq. \eqref{eq:samp6} leads to the following expression relating the nodal clustering coefficient of the sampled network, $c_S(k)$, to that of the original network:
 
\begin{equation}\label{eq:samp8}
c_S(k)=\frac{1-T}{1-T\langle c \rangle}\frac{P(k)}{P_S(k)}c(k)
\end{equation}

\noindent where $\langle c \rangle=\sum_kc(k)P(k)$. We have taken into account that the number of nodes of the sampled network is given by $N_S=(1-T\langle c \rangle)N$.

The edge clustering coefficient, $c(k,k^{\prime})$, which corresponds to the probability of an edge joining two nodes, one of degree $k$ and the other of degree $k^{\prime}$, share a common neighbor \cite{serrano2006a}. In other words, it can be interpreted as the probability of a link joining these two nodes to be the edge of a triangle. It is defined as $c(k,k^{\prime})=m_{k,k^{\prime}}/m^c_{k,k^{\prime}}$, where, $m_{k,k^{\prime}}$ is the average multiplicity of the edges linking degree classes $k$ and $k^{\prime}$ and $m^c_{k,k^{\prime}}=\mbox{max}(k,k^{\prime})-1$ is its maximum value.

Serrano \& Bogu\~n\'a \cite{serrano2006a} have shown that networks can only be considered uncorrelated when clustering is weak. In this case $m_{k,k^{\prime}}\simeq m_0$ with $m_0<1$ independent of $k$ and $k^{\prime}$, and therefore $c(k,k^{\prime})=m_0/m^c_{k,k^{\prime}}$. The corresponding edge clustering coefficient on the sampled network will therefore be given by  $c_S(k,k^{\prime})=m_S/m^c_{k,k^{\prime}}$ where $c_S$ is a constant to be determined in terms of $T$, $c_0$ and $\alpha$ where $c(k)=c_0/(k-1)^{\alpha}$. To do so this we use the following equation relating $m_{k,k^{\prime}}$ and $c_S(k)$ \cite{serrano2006b}:

\begin{equation}\label{eq:samp10}
\sum_{k^{\prime}>1}m_{k,k^{\prime}}P_S(k,k^{\prime})=\frac{k(k-1)}{\langle k \rangle}P_S(k)c_S(k)
\end{equation}

\noindent where $P_S(k,k^{\prime})=kP_S(k)P_S(k^{\prime}\vert k)/\langle k \rangle$ with $P_S(k^{\prime}\vert k)$. Now, by summing up $\forall k>1$, and using Eq. \eqref{eq:samp8}, we have

\begin{equation}\label{eq:samp11}
m_S=c_0\frac{1-T}{1-T\langle c \rangle}\frac{\langle k\rangle_U}{\langle k \rangle}\frac{\sum_{k>1}\frac{k}{(k-1)^{\alpha-1}}\frac{P(1)}{\langle k \rangle_U}}{1-2\frac{P_S(1)}{\langle k \rangle}+P_S(1,1)}
\end{equation}

\noindent where $m_S$ is the average edge multiplicity on the sampled networks and $\langle k\rangle$ and $\langle k\rangle_U$ is the average degree in the sampled and un-sampled networks, respectively.

The parameter $P_S(1,1)$ appearing in Eq. \eqref{eq:samp11} is given by \cite{alarcon2009}:

\begin{eqnarray}\label{eq:samp14}
\nonumber &&  P_S(1,1)=\frac{N\langle k \rangle_U}{N_S\langle k \rangle}\left(P(1,1)+\Sigma_{(1,n)\rightarrow (1,1)}+\Sigma_{(m,n)\rightarrow (1,1)}\right)\\
\nonumber && \Sigma_{(1,n)\rightarrow (1,1)}=p^2{\cal A}_{k=1}\sum_{n>1}\frac{\frac{P(1)}{\langle k \rangle_U}\frac{nP(n)}{\langle k \rangle_U}(1-p)^{n-1}}{\left(1-\frac{P(1)}{\langle k \rangle_U}\right)}\\
\nonumber && \Sigma_{(m,n)\rightarrow (1,1)}=p^2{\cal A}_{k>1}\left[\sum_{m>1}\frac{m\frac{P(m)}{\langle k \rangle_U}(1-p)^{m-1}}{\left(1-\frac{P(1)}{\langle k \rangle_U}\right)}\right]^2\\
\end{eqnarray}

\noindent where ${\cal A}_{k=1}=1-\langle k \rangle P(1,1)/P(1)$ and ${\cal A}_{k>1}=1-2P(1)/\langle k \rangle+P(1,1)$ \cite{serrano2006b}.

At this point we are ready to study the percolation process on the sampled network by applying the conditions derived in \cite{serrano2006b} for weakly clustered networks. 

\begin{eqnarray}\label{eq:res1}
\frac{\langle k(k-1)\rangle}{\langle k\rangle}\geq \left(1-\frac{P(1)}{\langle k \rangle}\right)\left(m_s+\frac{1-\frac{P_S(1)}{\langle k \rangle}}{1-2\frac{P_s(1)}{\langle k \rangle}+P_S(1,1)}\right)
\end{eqnarray}

Next we present results for two particular cases in which analytical results can be obtained \cite{alarcon2009}, namely, an exponential network with a degree distribution given by:

\begin{equation}\label{eq:exp1}
P(k)=(1-e^{-1/\beta})e^{-k/\beta},
\end{equation}

\noindent and a scale-free network characterized by a degree distribution given by \cite{newman2001}

\begin{equation}\label{eq:sf1}
P(k)=\frac{k^{-\alpha}e^{-k/\beta}}{\mbox{Li}_{\tau}(e^{-1/\beta})}.
\end{equation}

In both cases, we consider $c(k)=c_0/(k-1)^2$. Under these assumptions analytical, closed expressions for $G_S(x)$ can be obtained. 
   
\begin{figure}
\begin{center}
\includegraphics[scale=0.25]{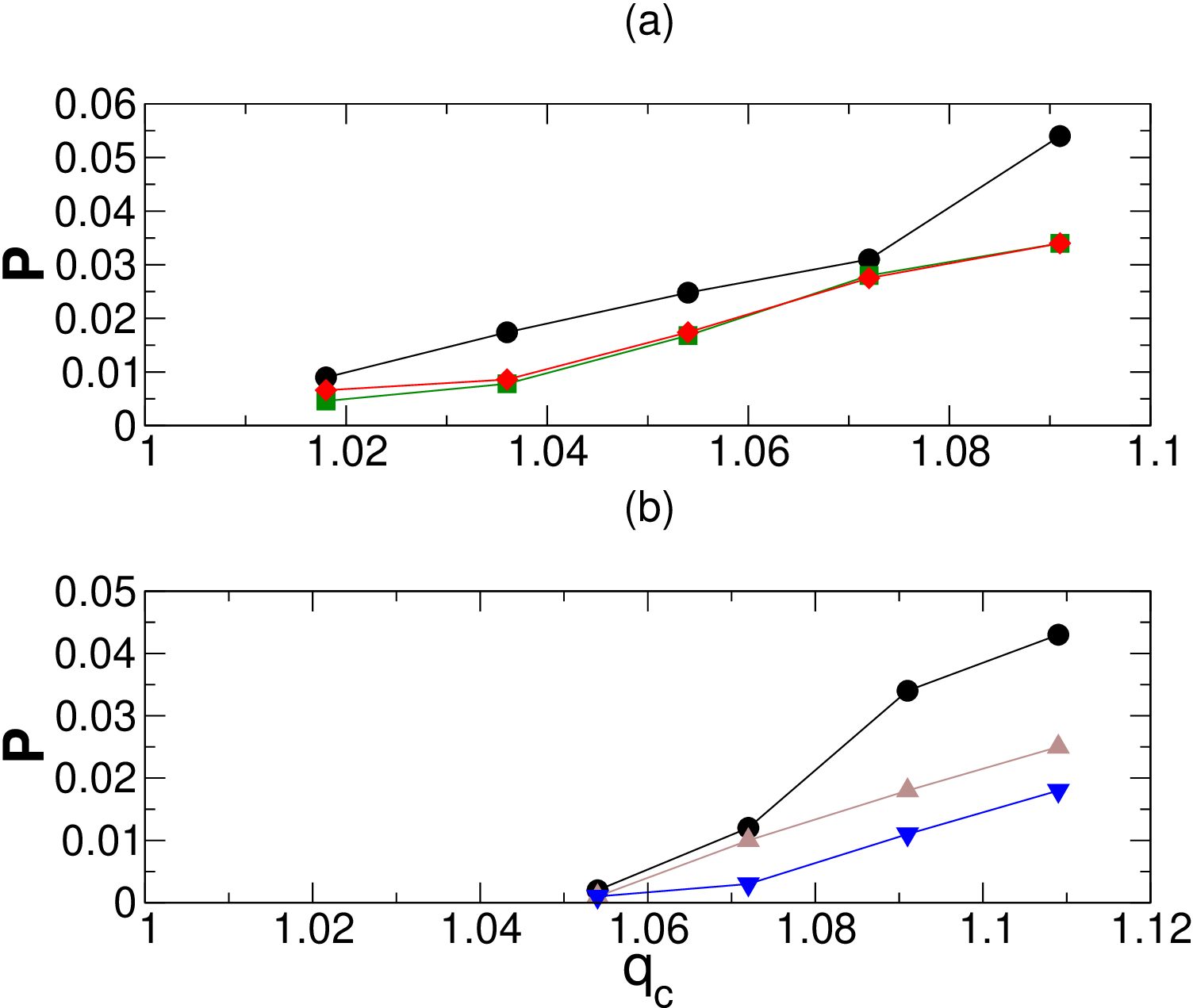}
\end{center}
\caption{Simulation results for weakly clustered networks with exponential degree distribution under sampling for the percolation probability ${\bf P}$ as a function of $q_c=\langle k(k-1) \rangle/\langle k \rangle$. In agreement with the analytical results shown in Fig. \ref{fig:5}, for lower clustering (panel (a) corresponding to $m=0.4$) ${\bf P}$ is bigger for $T=0.8$ than for $T=1$. In contrast, for larger clustering (panel (b) corresponding to $m=0.6$), ${\bf P}$ depends monotonously on $T$.  Color code: black lines (circles) correspond to the sampled network with $T=0$, brown lines (triangles pointing up) correspond to $T=0.4$, blue lines (triangles pointing down) correspond to $T=0.6$, green lines (squares) correspond to $T=0.8$, and red lines (diamonds) correspond to $T=1.$.\label{fig:3}}
\end{figure}

It is important to point out that the spreading process we consider is different from the percolation process on a network where the average multiplicity is reduced from $m_0$ to $(1-T)m_0$, i.e. to study percolation on a network with the same characteristics as the original network (same number of nodes, same number edges, etc), but with a given number of triangles removed. For this process, see \cite{serrano2006b}, the percolation threshold is reduced and, thus, percolation is facilitated.
The critical value of $\beta$ for which the equality sign in Eq. (\ref{eq:res1}) holds we denote by $\kappa$.   
 Fig. \ref{fig:5} shows that for a weakly clustered network, the percolation threshold, i.e. $\kappa$, on the restricted network is systematically bigger than the one corresponding to \emph{unrestricted} percolation and bigger than the corresponding percolation thresholds in networks with lower clustering (see Fig. \ref{fig:5}). This is mainly due to the effective removal of nodes and edges: although overall clustering is lowered, which, in principle, should favor the onset of percolation, at the same time nodes and edges are being removed, which acts against percolation. In spite of this overall trend, networks with an exponential degree distribution show a complex behavior under sampling. Fig. \ref{fig:5} demonstrates that the effect on percolation depends on both the average edge-multiplicity of the unsampled network, $m_0$, and the probability of sampling, $T$. We notice, that, for larger values of $m_0$, the onset of percolation is hindered as the network is more densely pruned (i.e. as $T$ increases). On the contrary, for smaller values of $m_0$, the onset of percolation does not depend monotonically on $T$: there exist a value of $T$ for which the corresponding percolation threshold reaches a maximum value and then starts decreasing, thus percolation being favored by further pruning. The reason why this behavior comes about is that, in the later case, a balance between network pruning and the related reduction in clustering is reached beyond which the decrease in clustering induced by further pruning outperforms the corresponding loss of connectivity. In the former cases, where clustering is stronger, this regime cannot be reached. These results are confirmed by computer simulations shown in Figs. \ref{fig:3}., where the probability of percolation, $\mathbf{P}$ is calculated for different values of $m_0$ and $T$. We observe that, for $m_0=0.4$ this quantity is bigger for $T=1$ than it is for $T=0.8$, whereas for $m_0=0.6$, $\mathbf{P}$ is a monotonically increasing function of $T$, in agreement with our analytic results. 

Let us now turn to how the effect of the sampling process depends on the nature of the degree distribution.
Comparing the left and right panel of Figs. \ref{fig:5}, it is clear that exponential and scale-free networks exhibit qualitative different behavior. In particular, we observe that the non-monotonic dependence of the percolation threshold on the parameter $T$, observed in the former case, is absent in the case of scale-free networks: even at lower values of the average multiplicity $m$ the percolation threshold in sampled scale-free networks increases monotonically with the proportions of triangles removed. This difference may be useful for probing of empirical networks and help to discern whether their degrees are distributed according to an exponential or to a scale-free distribution (with cut-off).

{\it Summary and discussion} -- We have studied spreading and percolation restricted to a part of a network. In our case we specifically avoid a certain fraction of triangles. For the case of weakly-clustered, uncorrelated networks we have given a full analytical description of the sampled networks in terms of the parameters of the original network. We have demonstrated  that  although edge and node removal may hinder percolation, the removal of triangles can lower the percolation threshold in situations where fewer triangles lead to less clustering. This effect depends on the functional form of the degree distribution and is found in exponential networks but not in scale-free networks. In scale-free networks we found that motif removal has a strong effect for moderate values of the average edge-multiplicity, in which case the onset of percolation increases dramatically with the removal of triangles. The reason for this is that with a finite cut-off, $\beta$, in the power law degree distribution low degree nodes dominate and are the nodes most likely to participate in the formation of triangles. This results in a largely disconnected network in which percolation is not possible. Percolation can only be obtained by  increasing $\beta$ as this allows for nodes with larger degrees which are less likely to form triangles and, therefore, to be removed from the network.

The strong dependence on the degree distribution of restricted spreading can presumably be  a useful way to probe the topological nature of big networks.
 
The authors would like to thank Renaud Lambiotte for valuable comments and suggestions. TA and HJJ gratefully acknowledge the EPSRC for funding under grant EP/D051223. TA would like to thank the Centre de Recerca Matem\`atica, Barcelona (Spain) for kind hospitatlity during the preparation of this manuscript.

\end{document}